\documentclass[prb,twocolumn,amsmath,amssymb,aps,floatfix,superscriptaddress]{revtex4-2}
\usepackage{graphicx}
\usepackage{amssymb}
\usepackage{amsmath}
\usepackage{stix}
\usepackage{physics}
\usepackage{xcolor}
\usepackage{bm}% bold math
\usepackage{hyperref}% add hypertext capabilities

\begin{document}

\title{Many-body skyrmion interactions in helimagnets}

%\title{Skyrmion-Skyrmion Interaction in Helimagnets}

\author{N. P. Vizarim}
\affiliation{Instituto de F\'isica "Gleb Wataghin", Universidade Estadual de Campinas - UNICAMP, 13083-859, Campinas, S\~ao Paulo, Brazil}

\author{J. C. Bellizotti Souza}
\affiliation{Instituto de F\'isica "Gleb Wataghin", Universidade Estadual de Campinas - UNICAMP, 13083-859, Campinas, S\~ao Paulo, Brazil}

\author{Raí M. Menezes}
\affiliation{Departamento de F\'isica, Universidade Federal de Pernambuco, Cidade Universit\'aria, 50670-901, Recife-PE, Brazil}

\author{Clécio C. de Souza Silva}
\affiliation{Departamento de F\'isica, Universidade Federal de Pernambuco, Cidade Universit\'aria, 50670-901, Recife-PE, Brazil}

\author{P. A. Venegas}
\affiliation{Departamento de F\'{i}sica, Faculdade de Ci\^{e}ncias, Universidade Estadual Paulista - UNESP, Bauru, SP, CP 473, 17033-360, Brazil}

\author{M. V. Milo\v{s}evi\'c}
\email{milorad.milosevic@uantwerpen.be}
\affiliation{COMMIT, Department of Physics, University of Antwerp, Groenenborgerlaan 171, B-2020 Antwerp, Belgium
}

\begin{abstract}
Many-body interactions strongly influence the structure, stability, and dynamics of condensed-matter systems, from atomic lattices to interacting quasi-particles such as superconducting vortices. Here, we investigate theoretically the pairwise and many-body interaction terms among skyrmions in helimagnets, considering both the ferromagnetic and conical spin backgrounds.
Using micromagnetic simulations, we separate the exchange, Dzyaloshinskii-Moriya, and Zeeman contributions to the skyrmion-skyrmion pair potential, and show that the binding energy of skyrmions within the conical phase depends strongly on the film thickness.
For small skyrmion clusters in the conical phase, three-body interactions make a substantial contribution to the cohesive energy, comparable to that of pairwise terms, while four-body terms become relevant only at small magnetic fields. As the system approaches the ferromagnetic phase, these higher-order contributions vanish, and the interactions become essentially pairwise. Our results indicate that realistic models of skyrmion interactions in helimagnets in the conical phase must incorporate many-body terms to accurately capture the behavior of skyrmion crystals and guide strategies for controlling skyrmion phases and dynamics.
\end{abstract}

\maketitle

\section{Introduction}

In magnetic materials lacking inversion symmetry, the handedness of the crystal structure, together with spin-orbit coupling, gives rise to an asymmetric exchange interaction known as the Dzyaloshinskii-Moriya (DM) interaction \cite{dzyaloshinsky_thermodynamic_1958,dzyaloshinsky_theory_1964}. This interaction stabilizes long-wavelength spatial modulations of magnetization with a uniform rotational orientation \cite{dzyaloshinsky_theory_1964,bak_theory_1980}. The past decade has witnessed a pertinently growing interest in such chiral helimagnetism, particularly following the discovery of localized non-collinear spin textures known as magnetic skyrmions \cite{bogdanov_thermodynamically_1988,bogdanov_thermodynamically_1994,rosler_spontaneous_2006,muhlbauer_skyrmion_2009,fert_magnetic_2017,romming_writing_2013,leonov_properties_2016}. Chiral interactions provide a unique stabilization mechanism that protects these localized states from instabilities \cite{bogdanov_thermodynamically_1994,leonov_properties_2016}, sparking significant interest in such systems for technological applications \cite{rosler_spontaneous_2006,li_chiral_2016,li_dynamics_2022}.

Magnetic skyrmions, in particular, are regarded as highly promising for future spintronic applications, including racetrack memories \cite{tomasello_strategy_2014,gobel_skyrmion_2021,juge_helium_2021,purnama_guided_2015}, logic operators \cite{luo_reconfigurable_2018,zhang_magnetic_2015,sisodia_programmable_2022}, neuromorphic computing \cite{li_magnetic_2017,chen_magnetic_2018,song_skyrmion-based_2020,yokouchi_pattern_2022}, and even quantum computing \cite{psaroudaki2021skyrmion}, among other applications \cite{finocchio2021magnetic}. Notably, skyrmions possess several advantageous characteristics, such as small size, stability, and potential for controlled motion, which can be manipulated through spin currents \cite{jonietz_spin_2010}, oscillating fields \cite{moon_skyrmion_2016}, and thermal or magnetic field gradients \cite{raimondo_temperature-gradient-driven_2022,zhang_manipulation_2018,chalus2025skyrmion}, making them highly versatile for future technological applications.

Despite the high promise, functional skyrmion-based devices remain scarce to date. A major challenge in realizing such devices lies in the precise control of skyrmion dynamics and stability~\cite{reichhardt_statics_2022}. Magnetic skyrmions can appear either as a lattice or as isolated particles within a magnetic material, and understanding the skyrmion–skyrmion interaction is a crucial step toward better control of these objects for practical applications. 

When embedded in the magnetically saturated phase \textemdash where all spins align parallel to each other and to the external field \textemdash skyrmions are well known to interact repulsively \cite{capic_skyrmionskyrmion_2020, lin_particle_2013, brearton_magnetic_2020} and tend to form a triangular lattice~\cite{muhlbauer_skyrmion_2009, munzer2010skyrmion, yu_real-space_2010}.   
%
% The skyrmion lattice (SkL) was first observed experimentally in the cubic transition metal compound MnSi~\cite{muhlbauer_skyrmion_2009}, and soon after, in several other three-dimensional helical magnets~\cite{munzer2010skyrmion} and thin films~\cite{yu_real-space_2010}.
%
However, the behavior of skyrmions in different magnetic phases, such as the conical phase~\cite{muhlbauer_skyrmion_2009}, introduces unique dynamics that remained less understood to date. Leonov \textit{et al.} \cite{leonov_three-dimensional_2016} mathematically investigated the stabilization
of various solitonic states within the conical phase of chiral ferromagnets,
which were later observed experimentally by Loudon \textit{et al.} \cite{loudon_direct_2018}.
Unlike skyrmions in the saturated ferromagnetic phase, which are axisymmetric, the three-dimensional skyrmions 
stabilized in the conical phase are inhomogeneous along their axes~\cite{menezes2024skyrmion}.
Additionally, skyrmions in the conical phase exhibit a mutually attractive interaction at intermediate ranges,
leading them to form clusters \cite{leonov_three-dimensional_2016,leonov_spintronics_2016}. 
This attractive interaction arises from the superposition of a broad,
strongly asymmetric transitional region dubbed \textit{shell} \cite{leonov_three-dimensional_2016}.
Further experimental work by Du \textit{et al.}
\cite{du_interaction_2018}
demonstrated that as the external magnetic field is increased, making the background more ferromagnetic, 
the interaction profile between skyrmions shifts from attractive to repulsive. Although several studies addressed key aspects of skyrmion interactions in helimagnets, a number of features remained unclear, such as the effect of the film thickness, and many-body interactions in systems with multiple-to-many skyrmions.

Therefore, in this work we conduct micromagnetic simulations to systematically investigate pairwise and multi-body skyrmion-skyrmion interactions in helimagnets. We first characterize the pairwise interactions of skyrmions as a function of relevant parameters. Specifically, we detail the dependence of skyrmion-skyrmion interactions on the applied magnetic field, showing that it can be tuned from attractive to purely repulsive, consistent with experimental observations \cite{du_interaction_2018}.
We further explore the impact of film thickness on the (relative) strength of attractive interactions, revealing that in thinner films the attractive interaction is weaker, while in thicker films the interaction becomes stronger, eventually reaching an asymptotic value corresponding to the bulk limit. Additionally, we demonstrate that at low applied magnetic field, multi-body contributions play a significant role in skyrmion-skyrmion interactions, as the interaction among multiple skyrmions differs from the simple sum of pairwise interactions. At high fields, when the applied field approaches (but remains below) the saturation field, the multi-body interactions vanish, and the interaction energy can be well approximated by the sum of pairwise contributions.

% Understanding both pairwise and multi-body skyrmion-skyrmion interactions is essential for the practical implementation of skyrmion-based devices. The ability to control these interactions could unlock new possibilities in data storage, logic devices, and neuromorphic computing. Therefore, we expect our results to provide significant insights for future technological applications.

The paper is organized as follows. In Section \ref{micromagnetic}, we present the micromagnetic model for the ferromagnetic film, considering exchange, Dzyaloshinskii–Moriya, and Zeeman energies; Section \ref{phase_diagram} presents a phase diagram that identifies the stable topological phases and the regions where skyrmions are stable; Section \ref{pairwise_interaction} shows the skyrmion–skyrmion interactions within the conical and saturated phases; Section \ref{three_body} demonstrates that three-body interactions are relevant for interacting skyrmions in the conical phase; Section \ref{multi_body} shows that higher-order multibody skyrmion interactions become relevant at low applied fields; and Section \ref{conclusions} provides final remarks and conclusions.

\section{THEORETICAL FRAMEWORK}

\subsection{Micromagnetic model}
\label{micromagnetic}

For the simulations involving skyrmions embedded within the conical phase,
we employ the micromagnetic simulator Mumax$^3$~\cite{vansteenkiste_mumax_2011,leliaert_fast_2018}. We consider a ferromagnetic film with dimensions of $1024\times 256\times d$~nm$^3$, where $d$ is the thickness of the film. Periodic boundary conditions are applied along the $x$ and $y$
directions, while open boundary conditions are imposed along the 
$z$ axis. The magnetic states are modeled by assuming the magnetic energy density functional

\begin{equation}\label{eq1}
    \mathcal{E}[\mathbf{m}] =  
    A_{ex}(\grad \mathbf{m})^2 + 
    D\mathbf{m} \cdot (\grad \times \mathbf{m}) -
    M_s \mathbf{m} \cdot \mathbf{H} \ ,
\end{equation}
where $\mathbf{m} = (m_x,m_y,m_z)$ is the normalized magnetization function. The first term on the right side of Eq.~\eqref{eq1} represents the exchange energy, 
the second term is the Dzyaloshinskii-Moriya coupling energy and 
the third term is the Zeeman energy, with $A_{ex}$ the exchange stiffness; $D$ the Dzyaloshinskii-Moriya coupling constant; $\mathbf{H}$ the applied magnetic field, and $M_s$ is the saturation magnetization.

For the simulations, we consider parameters $M_s = 3.84 \times 10^{5}$ A/m, $A_{ex} = 4.75 \times 10^{-12}$ J/m, and $D = 0.853 \times 10^{-3}$ J/m², stemming from B20-type FeGe \cite{zheng_experimental_2018}. The external field is applied perpendicular to the film, i.e., $\mathbf{H} = (0,0,H)$. The system is discretized into micromagnetic cells of size $2 \times 2 \times 2$ nm$^3$. Although intrinsic exchange and cubic anisotropies~\cite{bak_theory_1980} may define a preferential direction for spin rotation in helimagnets at zero field, such higher-order contributions are much weaker than the energy terms in Eq.~\eqref{eq1} and are therefore neglected in the calculations. Similarly, the small contribution from dipolar interactions in such materials is also not included~\cite{Maleyev_2006,Grigoriev_2006}, which particularly saves significant computational time in our three-dimensional simulations.

The dynamics of the magnetization is governed by the Landau-Lifshitz-Gilbert (LLG) equation
\begin{equation}\label{eq2}
    \dot{\bf{m}} = - \gamma \bf{m} \times \bf{H}_{\text{eff}} + \alpha \bf{m}\times\dot{\bf{m}},
\end{equation}
where $\gamma$ is the gyromagnetic ratio, $\alpha$ the dimensionless damping factor, and $\bf{H}_{\text{eff}}$ the effective field, which can be derived from the free energy~$E[\textbf{m}]=\int\mathcal{E}dV$ by taking the functional derivative with respect to the magnetization: $\bf{H}_{\text{eff}} = - \frac{1}{M_s}\delta \mathcal{E}/\delta \bf{m}$. In all simulations we took
$\alpha=0.3$.

% , and we consider
% $J=1$ meV and $D=0.18J$.
% The numerical integration of Eq.~\eqref{eq2}
% is performed using a fourth order Runge-Kutta
% method.

% A conical state is achieved in the film for certain values of $H$ where $L_D$ is the helix period. We use a cubic distribution of spin sites with distance between neighboring spins as $a = 2$ nm. 
% The helix period stabilized in our simulations is $L_D = 32a$.

\section{Results}
\label{section3}

\begin{figure}[t]
\centering
\includegraphics[width=0.9\linewidth]{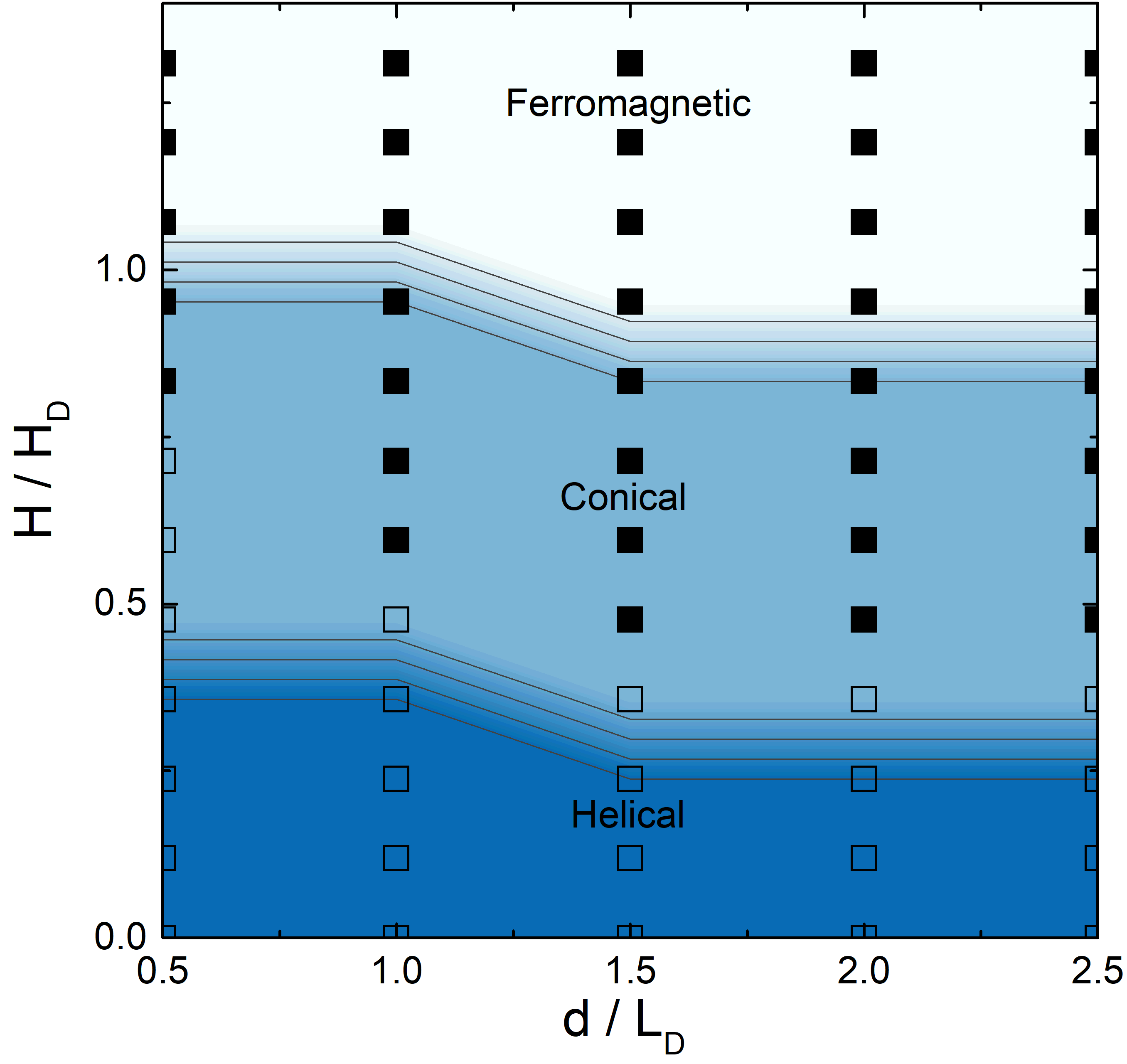}
\caption{
Magnetic field -- sample thickness phase diagram for the films simulated in this work. 
Different stable topological phases are indicated by shades of color. The black squares
correspond to attempts to stabilize skyrmions in the sample: solid squares mark parameter values
where skyrmions were stabilized and open squares mark values where skyrmions were not found
stable.
}
    \label{fig2}
\end{figure}

\subsection{Phase diagram}
\label{phase_diagram}

Before investigating the skyrmion-skyrmion interaction, we first examine the magnetic phases in the sample by varying the applied magnetic field \( H \) and the sample thickness \( d \), using the micromagnetic model described in Section~\ref{micromagnetic}. Figure~\ref{fig2} presents the resulting phase diagram, where the sample thickness and applied field are respectively normalized by \( L_D \), the period of the helical phase at zero field, and \( H_D \), the saturation field for the case \( d/L_D = 1 \).
The phase regions are depicted in shades of blue:
dark blue indicates the helical phase, light blue the conical phase, and white the ferromagnetic phase.
This phase diagram closely resembles those reported for FeGe/Si(111) epitaxial films \cite{ahmed_chiral_2018}.
Additionally, we attempted to stabilize skyrmions for selected values of $H/H_D$ and $d/L_D$, 
as indicated by the black squares in the figure. Solid black squares denote conditions where skyrmions
were successfully stabilized, while open squares represent conditions
where skyrmions could not be stabilized. This step is crucial for our investigation
of skyrmion-skyrmion interactions, as it is important to first identify the conditions
under which skyrmions remain stable or metastable.

\subsection{Pairwise skyrmion-skyrmion interaction}
\label{pairwise_interaction}

\begin{figure}[t]
\centering
\includegraphics[width=\linewidth]{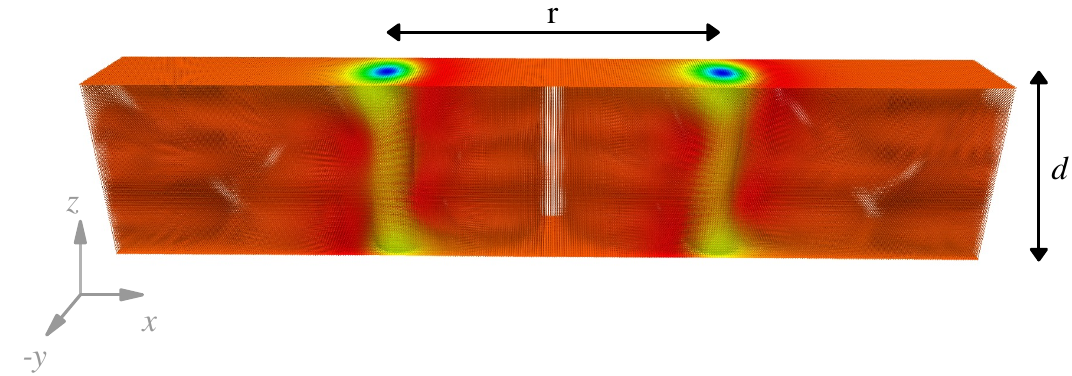}
\caption{
A 3D visualization snapshot of a spin configuration for the simulated ferromagnetic film of thickness $d$ and approaching
skyrmions at a distance $r$. 
The spin colors represent the $m_z$ spin-component, ranging from blue ($m_z=-1$) to red ($m_z=+1$).
}
    \label{fig1}
\end{figure}

We now turn our attention to the skyrmion-skyrmion interaction. 
As illustrated in the phase diagram in Fig.~\ref{fig2}, skyrmions can only be stabilized within a conical
or saturated ferromagnetic background. 

To calculate the skyrmion-skyrmion interactions in our simulations, two skyrmions are introduced into the system:
one at the center of the sample and the other at a distance $r$ along the
$x$ axis from the central skyrmion [see Fig.~\ref{fig1}]. To fix the distance between the considered skyrmions, we froze the spins at the centers of both skyrmions during the simulations.
The system is then relaxed, and the energy is minimized. This process is repeated for several sequential values of $r$, gradually bringing the skyrmions closer together to obtain a smooth interaction profile. This approach is similar to the method used in Refs.~\cite{menezes2019manipulation,muller2015capturing,lin2013particle,mulkers2017effects,stosic_pinning_2017} to calculate the interaction of the skyrmion with holes, sample edges, material defects and superconducting vortices.

\begin{figure}[t]
\centering
\includegraphics[width=\linewidth]{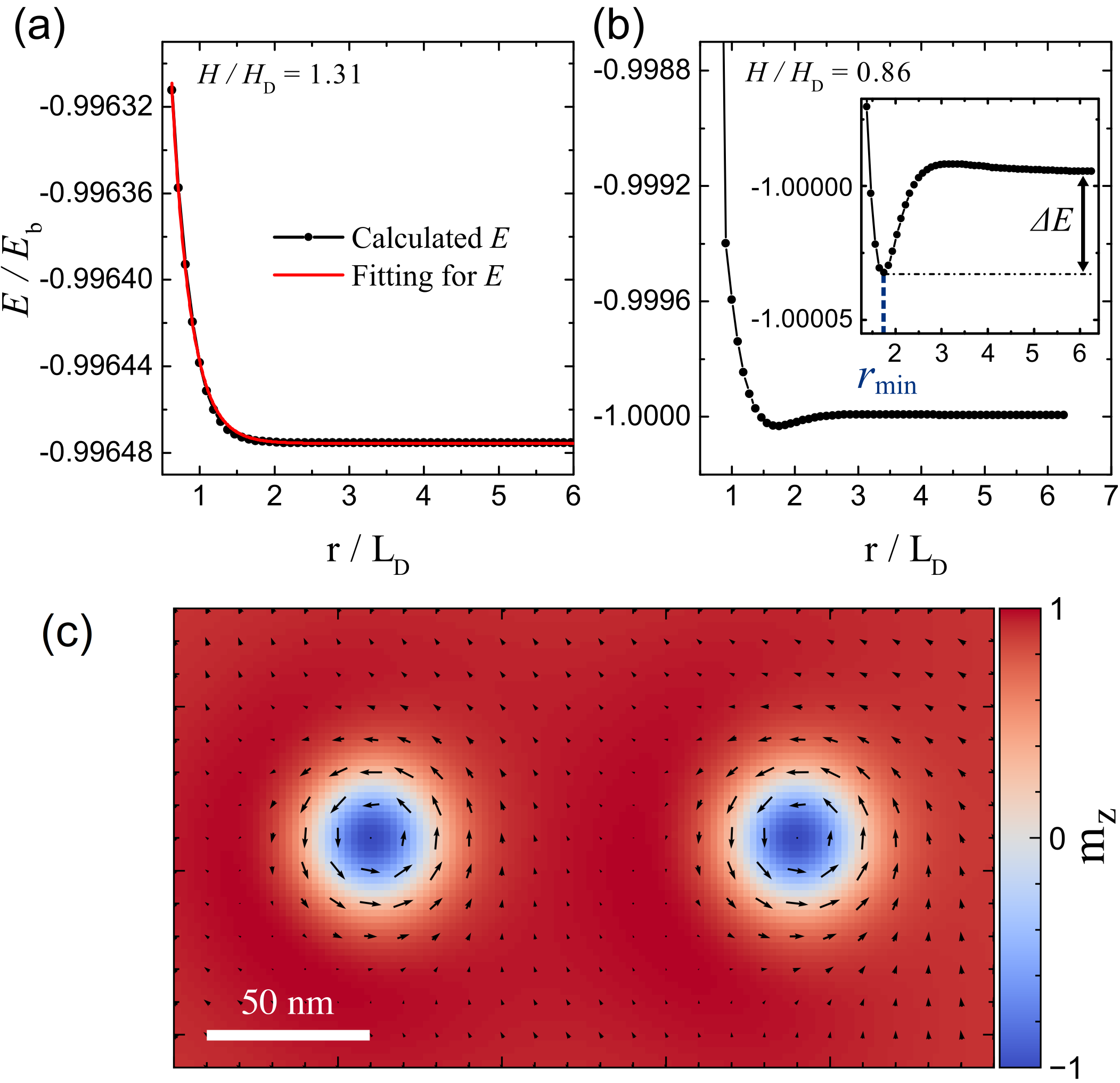}
\caption{
(a,b) Calculated energy of two skyrmions, normalized by the respective background energy $E_b$, i.e the energy of the system in the absence of skyrmions, as a function of the distance between the (centers of) skyrmions, $r/L_D$, for (a) $H/H_D=1.31$ (ferromagnetic background) and (b) $H/H_D=0.86$ (conical background). The solid red line in (a) shows the fitting function $E / E_{b} = a\exp(-r/L_Db) + c$, with
$a=0.00197$, $b=0.25280$ and $c=-0.99648$. The inset in (b) zooms on the attractive part of the interaction, where $r_{\mathrm{min}}$ is the distance corresponding to the minimal energy and $\Delta E$ is the depth of the attractive potential well.
(c) Spin configuration of two skyrmions in the conical phase for $H/H_D = 0.86$ and $r=2.125L_D=136$~nm.
In all cases, the considered film thickness was $d/L_D=1$.
}
    \label{fig3}
\end{figure}

To examine the skyrmion-skyrmion interaction in the saturated ferromagnetic phase, we set $d/L_D = 1$ and calculate the interaction energy profile for $H/H_D = 1.31$. 
The skyrmions are initially positioned at a distance $r_0=6.25L_D=400$~nm apart, 
and then gradually brought closer together.
As they approach, their energy is calculated using Eq.~\eqref{eq1}. The total energy $E$ increases, resulting in a monotonically repulsive interaction between the skyrmions, as depicted in Fig.~\ref{fig3}(a).
The energy profile was fitted as an exponential of the form $E / E_{b} \propto \exp(-r/bL_D)$, where $b$ is a fitting constant and $E_{b}$ the background energy, i.e., the energy of the system without skyrmions.
This finding aligns with previous studies on pairwise skyrmion interactions
in the saturated phase \cite{brearton_magnetic_2020,capic_skyrmionskyrmion_2020}.

\begin{figure*}[t]
\centering
\includegraphics[width=0.45\linewidth]{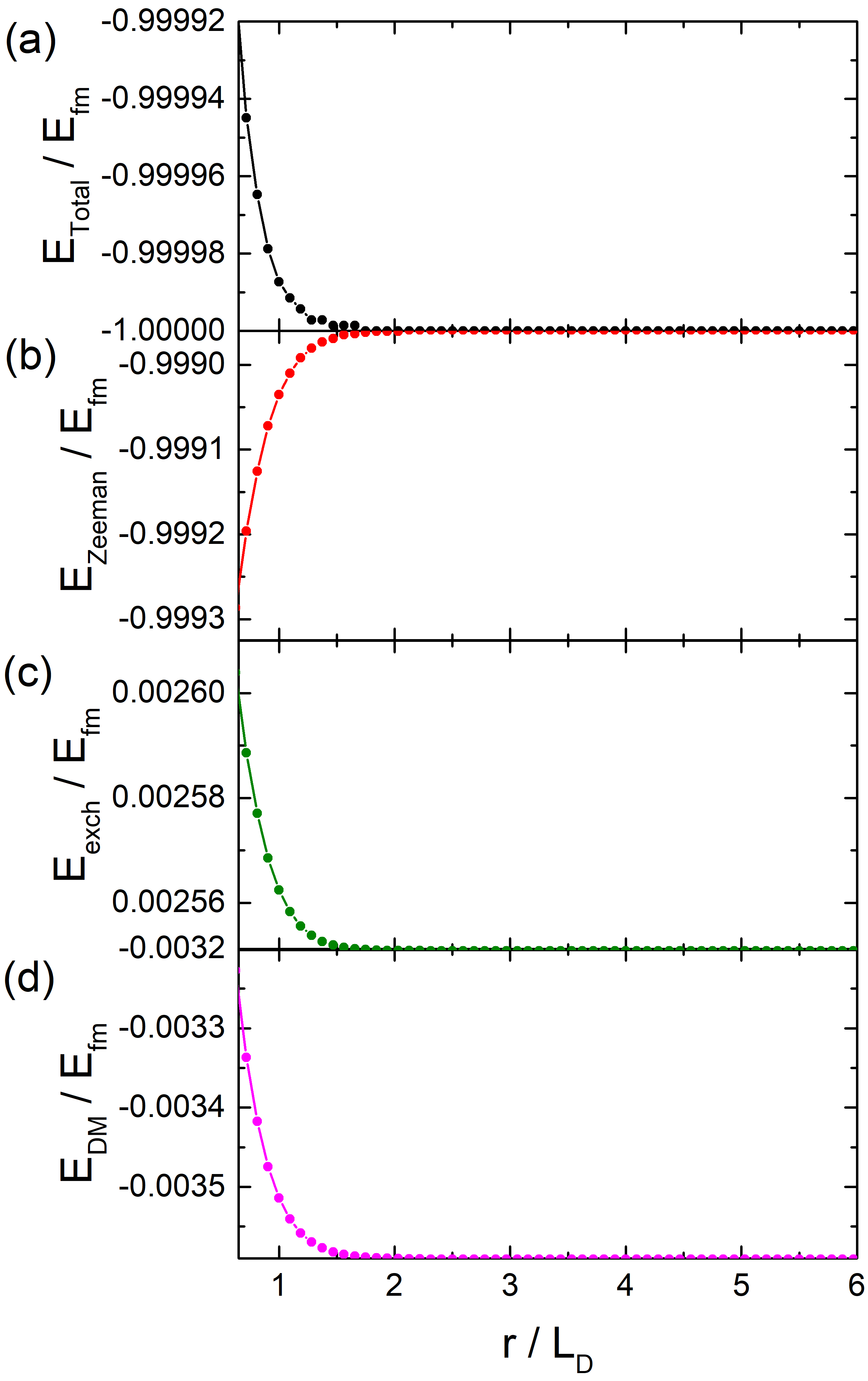}
\includegraphics[width=0.45\linewidth]{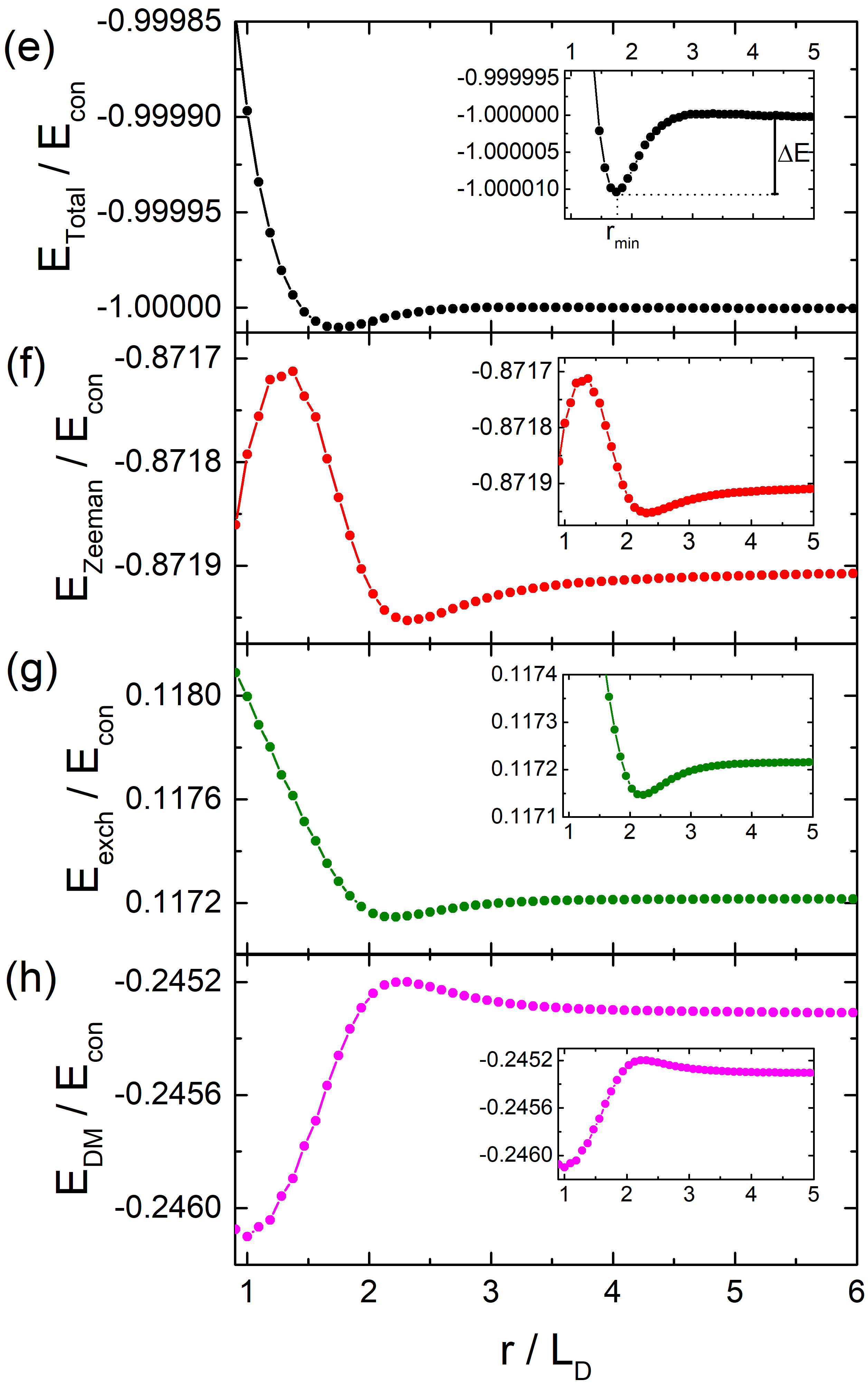}
\caption{
Total energy and its components (exchange, DM and Zeeman energies) of the two skyrmions as a function of the distance $r$ between them, in films with $d = 1.0L_D$, for (a–d) $H/H_D = 1.31$ and (e–h) $H/H_D = 0.86$. The energies are normalized by the respective energy background: $E_{\text{fm}}$ (saturated ferromagnetic) or $E_{\text{con}}$ (conical). The insets in (e–h) provide a zoomed-in view of the non-monotonic energy profiles, where $r_{\mathrm{min}}$ denotes the skyrmion separation at which the energy reaches a minimum, and $\Delta E$ represents the depth of the potential energy well.
}
    \label{fig4}
\end{figure*}

In contrast, when skyrmions are stabilized in the conical phase, the pairwise interaction exhibits significant changes.
As the skyrmions approach each other, they initially exhibit a slight repulsive interaction,
followed by a deep attractive potential well with a depth $\Delta E$, as shown in Fig.~\ref{fig3}(b).
This attractive interaction has an optimal distance $r_{\mathrm{min}}$, where the interaction energy reaches a minimum.
Upon further approach, the interaction becomes strongly repulsive.
The skyrmion-skyrmion interaction in the conical phase qualitatively resembles the shape of a Lennard-Jones potential \cite{lennard-jones_cohesion_1931}.
% Figure \ref{fig4} shows the total energy 
% $E$ relative to the corresponding background energy: ferromagnetic %$E_{fm}$ in Fig.~\ref{fig4}(a) 
%and conical $E_{con}$ in Fig.~\ref{fig4}(b).
It is noteworthy that this skyrmion-skyrmion interaction in the conical phase has already been described by Leonov \textit{et al.} \cite{leonov_three-dimensional_2016}, 
where the origin of the attractive interaction is attributed to the overlap of magnetization \textit{shells} 
that envelop the non-axisymmetric skyrmions in the conical phase.

Fig.~\ref{fig4} presents the total energy and its contributions for a film thickness of $d/L_D = 1$ in two scenarios:
(i) when the system is in the saturated ferromagnetic phase, under field $H/H_D = 1.31$, as shown in Figs.~\ref{fig4}(a–d); and
(ii) when the system is in the conical phase, with $H/H_D = 0.86$, as shown in Figs.~\ref{fig4}(e–h).

As expected, for $H/H_D > 1$, the skyrmion-skyrmion interaction is monotonically repulsive. The individual energy components [Figs. ~\ref{fig4}(b-d)] reveal that this repulsive interaction originates from the DM and exchange contributions, while the Zeeman energy favors an attractive interaction. In contrast, for $H/H_D < 1$, the spins in the sample do not align perfectly with the external field, leading to the formation of a conical phase. In this case, the skyrmion-skyrmion interaction becomes more complex and non-monotonic, as depicted in Figs. ~\ref{fig3} (b) and ~\ref{fig4}(e). 
As shown in Figs.~\ref{fig4}(f–h), the repulsive interaction at large distances is primarily driven by the DM interaction, while the exchange and Zeeman contributions remain attractive. In contrast, at short distances, the repulsive interaction originates from the exchange and Zeeman contributions, as the DM contribution becomes attractive.
At very short separations, $r/L_D < 1.5$, the Zeeman contribution becomes attractive again, similar to the ferromagnetic case shown in Fig.~\ref{fig4}(b). Even so, the total energy remains repulsive because the strongly repulsive exchange term dominates.

\subsubsection{Varying the applied field for fixed sample thickness}

Here we further explore the behavior of the non-monotonic skyrmion-skyrmion interaction by varying key conditions. First, we examine the influence of the applied magnetic field
while keeping the sample thickness constant at $d/L_D = 1$. 

\begin{figure}[t]
\centering
\includegraphics[width=\linewidth]{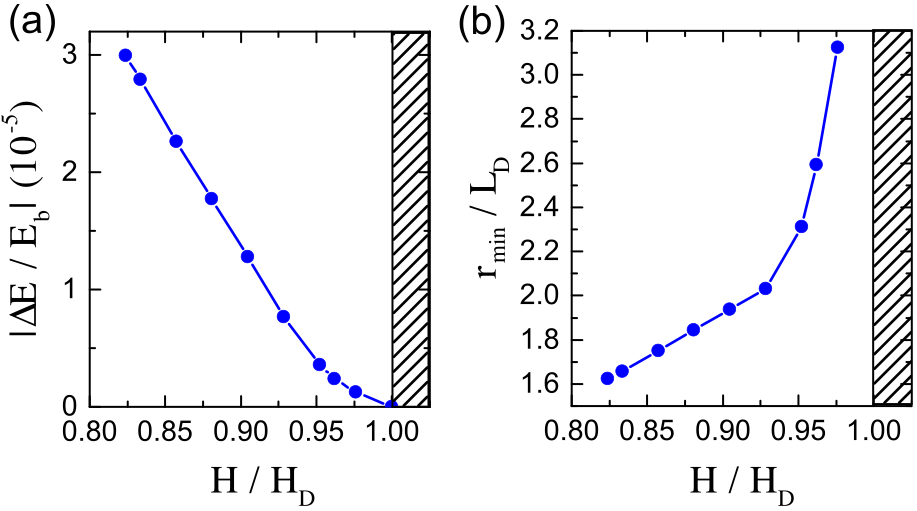}
\caption{
(a) The optimal distance, $r_{\mathrm{min}}/L_D$, and (b) the attractive potential depth, $|\Delta E/E_{b}|$, 
as a function of the applied field, $H/H_D$, for a sample of thickness $d=1.0L_D$. The dashed region indicates the saturated ferromagnetic phase.
}
    \label{fig5}
\end{figure}

Fig.~\ref{fig5}(a) shows how the attractive depth $|\Delta E/E_{b}|$
changes as a function of the applied field. There is an almost linear reduction in $|\Delta E/E_{b}|$
as $H/H_D$ increases. When $H/H_D$ reaches unity, the background becomes the ferromagnetic state and 
$\Delta E = 0$, indicating a transition of the skyrmion-skyrmion interaction from non-monotonic to monotonically repulsive, as shown in Fig.~\ref{fig4}(a). Fig.~\ref{fig5}(b) illustrates how the optimal distance $r_{\mathrm{min}}$ varies with the applied magnetic field $H/H_D$. At low and intermediate field strengths, $r_{\mathrm{min}}$ shows an almost linear dependence on the applied field.
However, at higher fields, $r_{\mathrm{min}}$ increases rapidly toward infinity.
This behavior closely resembles results obtained by Du \textit{et al.} \cite{du_interaction_2018}, 
where the influence of the magnetic field
on optimal skyrmion-skyrmion distance in the conical phase was experimentally investigated. At lower values of $H/H_D$, the background remains conical, and the attractive component of the
skyrmion-skyrmion interaction is stronger, leading to a shorter optimal distance between the skyrmions.
As $H/H_D$ increases, the background becomes increasingly ferromagnetic, 
reducing the attractive interaction and resulting in a more distant optimal skyrmion separation.
In the extreme case where the background is fully saturated, the attractive interaction disappears,
and the optimal distance approaches infinity, that is, skyrmions stabilize as far apart as possible.
When the number of skyrmions does not change and a lattice has formed, weaker fields tend to compress the cluster, whereas stronger fields loosen it. By contrast, during nucleation from the conical phase, the inter-skyrmion distance remains roughly constant both within clusters and in the resulting lattice \cite{kim_mechanisms_2020}.

\begin{figure}[t]
\centering
\includegraphics[width=\linewidth]{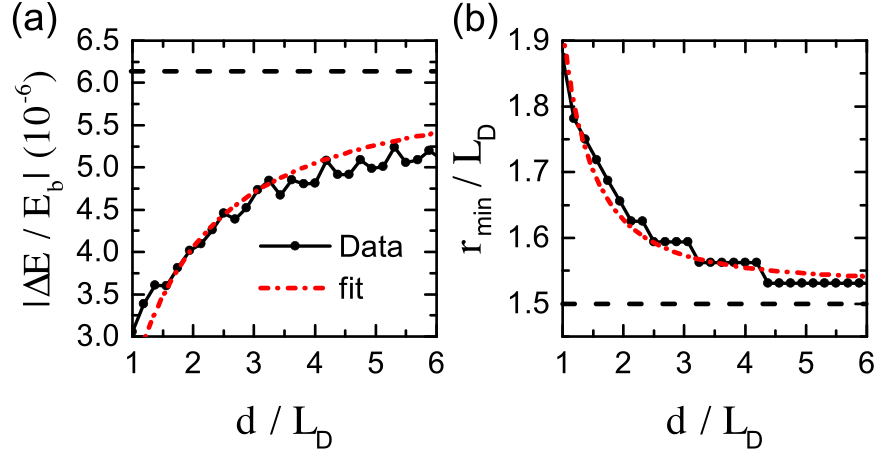}
\caption{
(a) The attractive potential depth, $|\Delta E/E_{b}|$, and (b) the optimal skyrmion separation, $r_{\mathrm{min}}/L_D$, 
as a function of the sample thickness, $d$, for a fixed applied field, $H/H_D = 0.77$. In both figures, the red dash-dotted curves correspond to approximated behaviors $|\Delta E/E_{b}|=\Delta E_{bulk} [1 - 1/\sqrt{1 + (d/\lambda)^2}]$ in (a) and $r_{\mathrm{min}}= \xi/d^2+r_{\mathrm{min}}^{bulk}$ in (b), where $\lambda$ and $\xi$ are fitting parameters. Black horizontal dashed lines correspond to the bulk limit of $\Delta E_{bulk}=6.14\times10^{-6}$ in (a), and
$r_{\mathrm{min}}^{bulk}=1.5L_D$ in (b).
}
    \label{fig6}
\end{figure}

\begin{figure*}
\centering
\includegraphics[width=\linewidth]{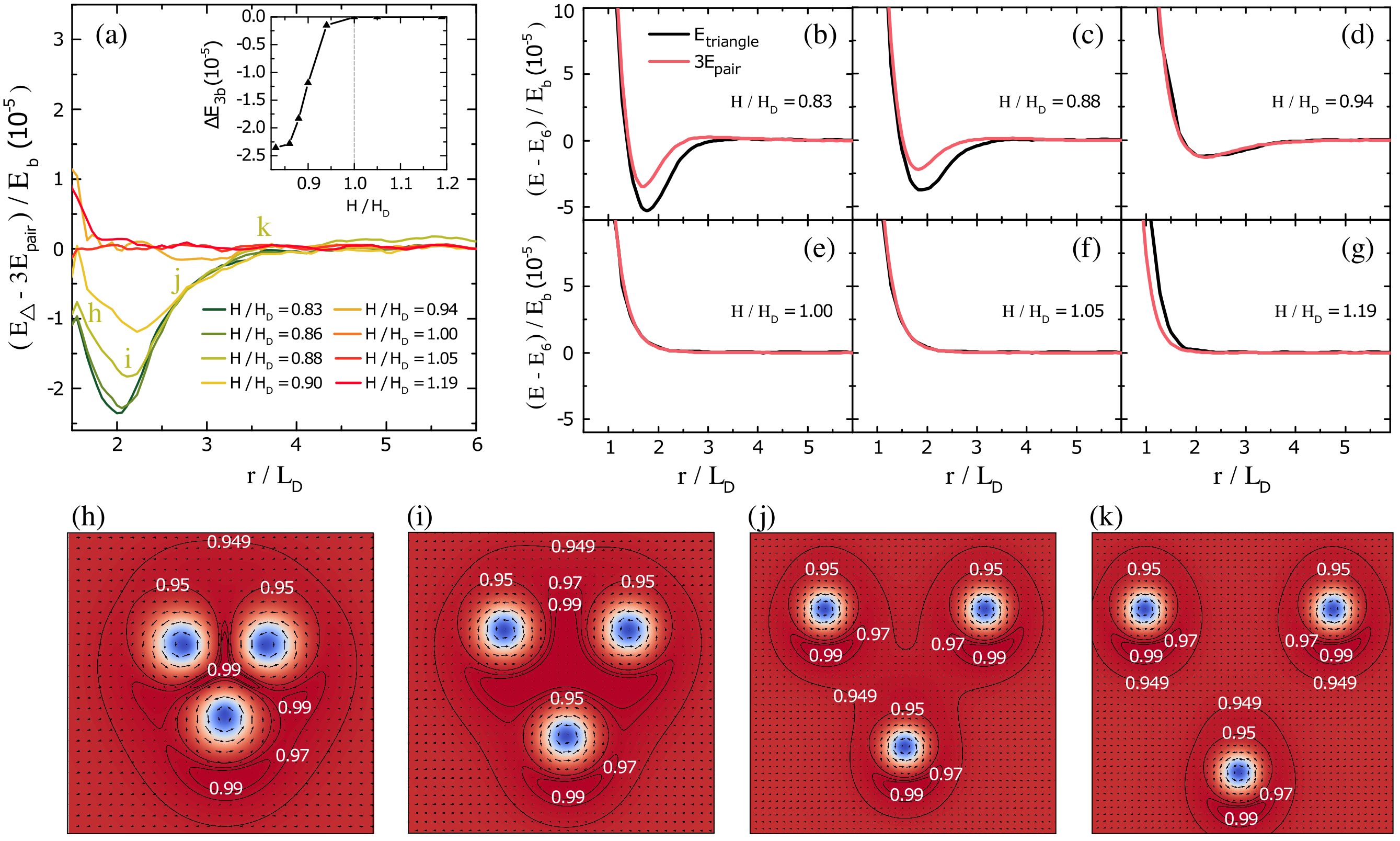}
\caption{
(a) Three-body skyrmion-skyrmion interaction energy profiles, defined as the energy difference \((E_{\triangle} - 3E_{\text{pair}})/E_b\), where \(E_b\) is the background energy of the sample, i.e in the absence of skyrmions, as a function of the distance between skyrmions \(r/L_D\), calculated for different values of the applied field. The inset shows the depth of the three-body potential well, \(\Delta E_{3b}\), as a function of the applied field. The letters label the scenarios where the magnetic configurations shown in (h–k) are observed. (b–g) Comparison between the energy of the three-skyrmion system (triangle of skyrmions) and the sum of the energies of isolated pairs, where the difference between the curves arises from the three-body contribution. Here $E_6$ is the energy calculated at $r/L_D=6$. (h–k) Top-view spin configurations of the three-skyrmion system in the conical phase for \(H/H_D = 0.88\), at different skyrmion separations: (h) \(r/L_D = 1.0\), (i) \(r/L_D = r_{\mathrm{min}} = 1.84\), (j) \(r/L_D = 2.96\), and (k) \(r/L_D = 3.5\). The colormap represents the \(z\)-component of the magnetization, consistent with Fig.~\ref{fig3}.
}
    \label{fig7}
\end{figure*}

\subsubsection{Varying the sample thickness for fixed applied field}

In this section, we investigate the skyrmion--skyrmion interaction for a fixed applied magnetic field, 
\( H/H_D = 0.77 \), while varying the sample thickness \( d \). 
As indicated in Fig.~\ref{fig2}, skyrmions can be stabilized in the conical phase at this field strength
and sample thickness. Notably, in thicker films with large \( d \) values, the increased number of micromagnetic spin cells leads to a higher propensity
for metastable configurations, hindering the energy minimization process and demanding significantly more computational time. For this reason, in this subsection we simulate a cross-section of the magnetic film, i.e., a cut in the \( x \)--\( z \) plane that passes through both skyrmion centers, effectively capturing the interaction of two adjacent chiral domain walls. Despite its limitations, this approach provides valuable qualitative insight into the analogous behavior of skyrmions. In this scenario, the energy profiles qualitatively reproduce the expected results of the full 3D simulations of two interacting skyrmions.

Figure~\ref{fig6}(a) illustrates how the depth of the attractive potential well, \( |\Delta E/E_{b}| \),
varies with \( d/L_D \). For very thin films, i.e., small values of \( d \), the skyrmion--skyrmion interaction is weaker 
compared to that in thicker films. 
As shown in Fig.~\ref{fig6}(a), the calculated data can be approximated by the expression $|\Delta E/E_{b}|=\Delta E_{bulk} [1 - 1/\sqrt{1 + (d/\lambda)^2}]$
where \( \Delta E_{\mathrm{bulk}} \) is the energy in the bulk limit and \( \lambda \) is a fitting parameter. 
This behavior arises because thicker films contain multiple layers of whirling spins in the conical state,
which enhances the attractive interaction between skyrmions.
In contrast, reduced thickness weakens the conical background due to surface effects, thereby suppressing the attractive interaction.
Note that, at large thicknesses, \( |\Delta E/E_{b}| \) tends to saturate at a fixed value corresponding to the bulk limit. In the bulk case, where there is no surface along the \( z \) direction, we identify an attractive potential depth of \( |\Delta E/E_{b}| = 6.14 \times 10^{-6} \) and an optimal skyrmion distance \( r_{\mathrm{min}} = 1.5 L_D \).

Fig.~\ref{fig6}(b) depicts the behavior of $r_{\mathrm{min}}$ as a function of the sample thickness. The results reveal that the optimal distance between skyrmions decreases slightly as the film thickness increases, being well described by $r_{\mathrm{min}} \propto 1/d^2$.

\subsection{Three-body skyrmion-skyrmion interaction}
\label{three_body}

Throughout this work, we have focused on pairs of interacting skyrmions in either the saturated ferromagnetic or the conical phase. However, it is also important to consider interactions involving multiple skyrmions, particularly in the conical phase, where attractive forces promote clustering.

In this section, we explore the interaction energy of a trio of skyrmions, arranged in an equilateral triangle. Similarly to the previous section, the spins are fixed at the centers of the skyrmions during the simulations, and the system energy is calculated as a function of the skyrmion separation (i.e., the side length of the triangle). The interaction among a trio of skyrmions exhibits a significantly stronger attractive force compared to that of three independent pairs. This enhanced attraction arises because more skyrmion \textit{shells} overlap and interact, strengthening the overall attractive potential. This indicates that the interaction among a trio of skyrmions is not merely the sum of three pairwise interactions, suggesting the presence of a significant three-body component in the system.

The three-body skyrmion interaction energy, defined as \( E_{3b} = E_{\triangle} - 3E_{\text{pair}} \), is shown in Fig.~\ref{fig7}(a) for different values of the applied magnetic field. Notably, \( E_{3b} \) decreases as the magnetic field increases, that is, as the background becomes more ferromagnetic.
%which corresponds to a more ferromagnetic background. 
The inset of Fig.~\ref{fig7}(a) shows the depth of the three-body potential well, \( \Delta E_{3b} \), as a function of the magnetic field, demonstrating a monotonic reduction. %in the interaction strength with increasing field.

Figures~\ref{fig7}(b--g) compare the energy profiles of the skyrmion trio with three times the energy profile of a pair of skyrmions, from which the curves in Fig.~\ref{fig7}(a) are obtained. At \( H/H_D = 0.94 \) [Fig.~\ref{fig7}(d)], slightly below the saturation field (\( H/H_D = 1.0 \)), the curves \( 3E_{\text{pair}} \) and \( E_{\triangle} \) approximately coincide, indicating that the three-body interaction vanishes. However, for \( H/H_D > 1.19 \), the curves separate again [see, e.g., Fig.~\ref{fig7}(g)], but only at short distances \( r/L_D < 2.0 \). We note that this might be an artifact of the frozen-core technique used to calculate the energy profiles, which becomes less reliable at short distances where skyrmions can be deformed. This implies that, in the high-field regime (ferromagnetic phase), where skyrmions interact repulsively, the three-body interaction might become relevant only in systems with a high skyrmion density, where skyrmions can be brought into close proximity.

\begin{figure}[t!]
\centering
\includegraphics[width=\linewidth]{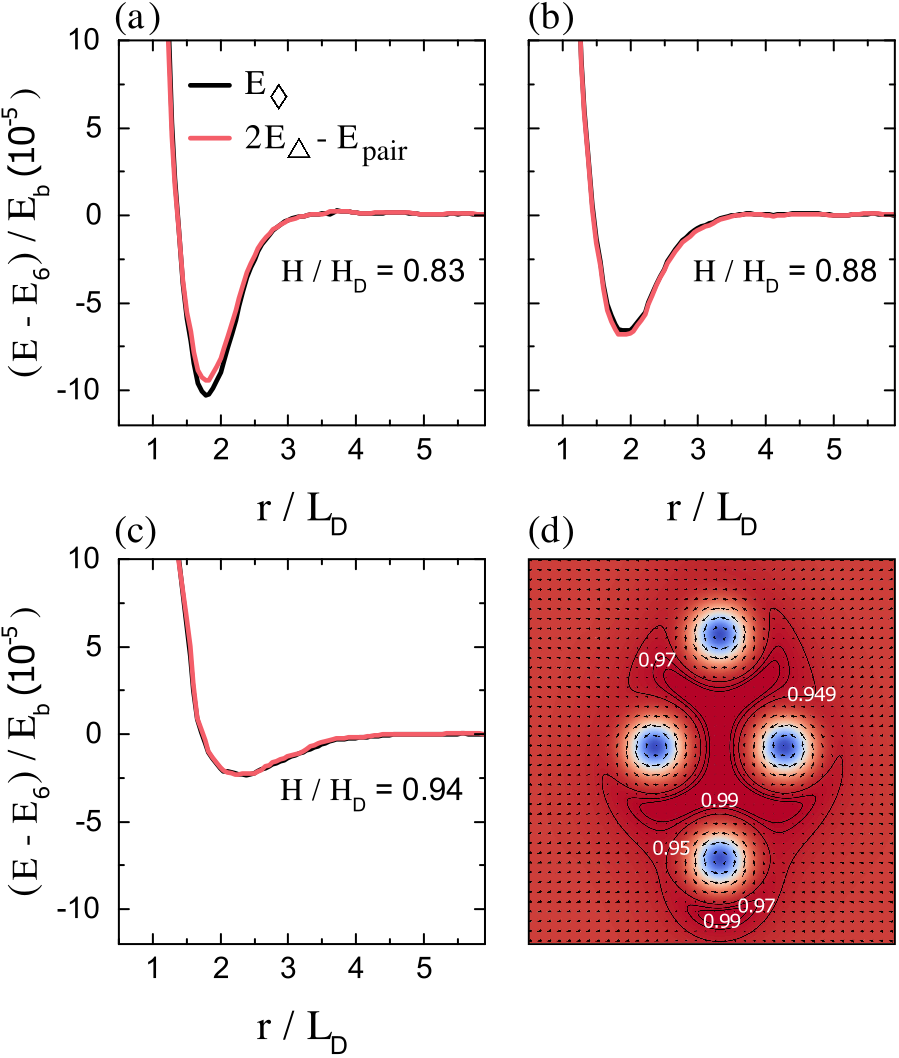}
\caption{
(a-c) Comparison of the energy profiles for different values of the applied magnetic field between a four-skyrmion cluster, arranged in a parallelogram ($E_\Diamond$), and a combination of two isolated three-skyrmion systems, each arranged in a triangular configuration ($E_\triangle$). In the latter case, the energy of a skyrmion pair, \( E_{\text{pair}} \), is subtracted to avoid double counting of one pairwise interaction. A difference between the curves is observed only for fields \( H/H_D < 0.88 \), where interactions beyond the nearest-neighbor pairwise and three-body interactions become non-negligible. Here $E_6$ is the energy calculated at $r/L_D=6$.
(d) Top view of the spin configuration of the four-skyrmion cluster in the conical phase for \( H/H_D = 0.83 \).
}
    \label{fig9}
\end{figure}

In comparison with the case of pairwise skyrmion interaction, the interaction between the \textit{shells} in a skyrmion trio is stronger, leading to a more attractive overall interaction, especially at lower magnetic fields. It is also noteworthy that the three-body interaction between skyrmions is non-monotonic. Typical three-body interactions in other condensed matter systems, such as vortices in superconductors~\cite{Chaves2011,Edstrom2013} and colloids~\cite{Brunner2004,Dobnikar2004,Correia2023}, are usually either monotonically repulsive or attractive. It is worth noting that type 1.5 superconducting vortices also exhibit nonmonotonic interactions \cite{moshchalkov_type-15_2009}. Here, in the skyrmion case, the nonmonotonicity arises from the changes in the magnetic state between skyrmions as they approach each other, similar to the pairwise interaction. 

Figures~\ref{fig7}(h)--(k) illustrate the central-layer spin configuration of the magnetic film containing a skyrmion trio under \( H/H_D = 0.88 \) for different skyrmion separations. When the skyrmions are far apart [e.g., Fig.~\ref{fig7}(k)], the magnetic state between them remains in the conical phase, and the interaction is attractive. As the skyrmions move closer to each other, the transitional rotated regions around them—the \textit{shells}—begin to overlap [Fig.~\ref{fig7}(j)], until a locally saturated magnetic state emerges between the skyrmions [Fig.~\ref{fig7}(i)]. Upon further reduction of the separation, the saturated region between skyrmions becomes compressed [Fig.~\ref{fig7}(h)], leading to an increase in energy. This indicates that the interaction becomes repulsive when a saturated ferromagnetic state forms between closely spaced skyrmions.

\subsection{Higher order multi-body skyrmion-skyrmion interaction}
\label{multi_body}

Last but not least, we investigate the role and magnitude of higher-order multi-body skyrmion interactions in the conical phase. To this end, we simulate a four-skyrmion cluster arranged in the diamond configuration [see Fig.~\ref{fig9}(d)], which is the smallest quartet found within a triangular skyrmion lattice and captures both nearest- and next-nearest-neighbor interactions within a single geometry.

Similarly to the previous section, to explore the interaction energy of the four-skyrmion system, the spins are fixed at the centers of the skyrmions during the simulations, and the system energy is calculated as a function of the skyrmion separation (i.e., the side length of the parallelogram). To isolate higher-order multi-body contributions, we compare the resulting energy profiles with the sum of the energies of two separate equilateral triangles (each analyzed in the previous section), subtracting the energy of the horizontally aligned skyrmion pair to avoid double counting. The results are shown in Fig.~\ref{fig9}(a)--(c) for different values of the applied magnetic field. A discrepancy between these two curves appears only for lower magnetic fields (\( H/H_D < 0.88 \)), indicating the presence of one or more of the following contributions: (i) four-body term, (ii) three-body terms from the nonequilateral trios, and (iii) next-nearest-neighbor pair interactions (skyrmions at the long diagonal of the diamond). For \( H/H_D \geq 0.88 \), the curves show excellent agreement, indicating that higher-order multi-body interactions, other than nearest-neighbor pairs and equilateral trios, are negligible in this regime.
%The results are shown in Fig.~\ref{fig9}(a)--(c) for different values of the applied magnetic field. Any discrepancy between these two curves indicates the presence of \ccss{four-body contributions and/or two-body interactions involving next-nearest neighbors higher-order interactions, such as four-body contributions or three-body interactions involving second-nearest neighbors.

%As seen in Fig.~\ref{fig9}(a)--(c), a small discrepancy between the curves appears only for lower magnetic fields (\( H/H_D < 0.88 \)). For \( H/H_D \geq 0.88 \), the curves show excellent agreement, indicating that higher-order multi-body interactions are negligible in this regime.

%\subsection{\ccss{Implications to the cohesive energy of the skyrmion crystal}}

\begin{figure}[t!]
\centering
\includegraphics[width=0.75\linewidth]{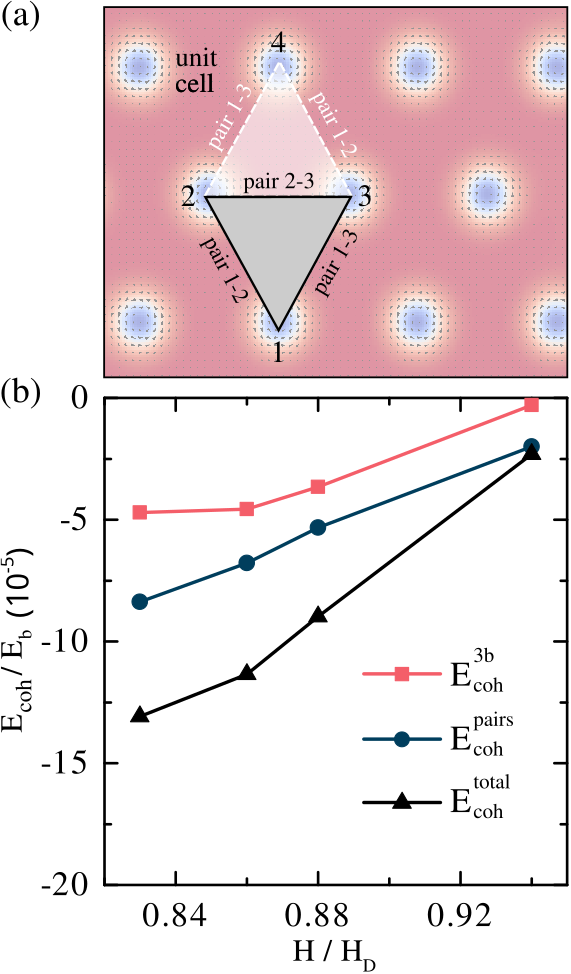}
\caption{ (a) Illustration of the spin configuration of a skyrmion crystal, highlighting the parallelogram unit cell.
(b) Cohesive energy density of the skyrmion crystal as a function of the applied magnetic field,
for films with $d = 1.0L_D$, approximated as $E_{\text{coh}}=E_{\text{coh}}^{\text{pairs}}+E_{\text{coh}}^{3b}$, where $E_{\text{coh}}^{\text{pairs}}=3\Delta E_{\text{pair}}$ (three distinct nearest-neighbor pairs per unit cell) and $E_{\text{coh}}^{3b}=2\Delta E_{3b}$ (two distinct triangles per unit cell) follows from a direct extrapolation of the isolated-diamond decomposition. In the lattice, the effective values of $\Delta E_{\text{pair}}$ and $\Delta E_{3b}$ may differ due to collective effects and periodic boundary conditions; the formula shown here is therefore illustrative rather than quantitative. }
    \label{fig10}
\end{figure}

Building upon the findings that higher-order terms beyond nearest-neighbor pairs and equilateral trios are negligible at moderate fields, the cohesive energy per unit cell of a triangular skyrmion lattice can be estimated as
\( E_{\text{coh}} = 2\Delta E_{3b} + 3\Delta E_{\text{pair}} \),
where the coefficients reflect the number of distinct nearest-neighbor pairs (three) and triangular triplets (two) in each diamond unit cell. Figure~\ref{fig10}(a) illustrates this geometry. For a qualitative estimate, we take the values $\Delta E_{\text{pair}}$ and $\Delta E_{3b}$ entering the above expression as those extracted from the single-diamond configuration; these will generally differ in an extended crystal due to the higher coordination, collective distortions of the skyrmion profiles, and the modified magnetostatic and exchange boundary conditions imposed by periodicity. The above relation should therefore be regarded as a qualitative guide to the hierarchy of interaction terms rather than a quantitative prediction~\footnote{A rigorous determination of $E_{\text{coh}}$ for the infinite lattice would require micromagnetic simulations with periodic boundary conditions, combined with energy minimization for varying cell sizes to allow the lattice to relax and to separate genuine many-body contributions from finite-cell artifacts.}.

Although approximate, this estimate still captures the relative weights of the pairwise and three-body terms, as shown in Fig.~\ref{fig10}(b). Interestingly, three-body interactions play a significant role in the cohesive energy of the skyrmion crystal, contributing almost as much as the pairwise interaction energy in the conical phase. This highlights the importance of including three-body interactions when modeling skyrmion statics and dynamics in helimagnets.

\section{Conclusions}
\label{conclusions}

In summary, we carried out a detailed investigation of pairwise and multi-body contributions to skyrmion interactions in helimagnets, for both fully saturated ferromagnetic and conical backgrounds. Our analysis of pairwise interactions extends previous work by Leonov \textit{et al.}~\cite{leonov_three-dimensional_2016} and Du \textit{et al.}~\cite{du_interaction_2018} by disentangling the contributions of exchange, Zeeman, and Dzyaloshinskii-Moriya energy terms, and by revealing the dependence of the skyrmion-skyrmion binding energy on the film thickness. These aspects, not addressed in earlier studies, provide a more detailed microscopic picture of skyrmion-skyrmion interactions, which could help guide strategies to control skyrmion stability in magnetic films. 

Exploring beyond pairwise interactions, our investigation of small skyrmion clusters revealed a prominent contribution of three-body terms for skyrmion trios in a conical background. This suggests that the asymmetric shell structure of skyrmions in the conical phase not only contributes to the attractive component of the pairwise interaction but also promotes higher-order contributions. As the external magnetic field increases and the conical state gives way to a fully saturated ferromagnetic background, these three-body contributions vanish. For skyrmion quartets, we find that an additional four-body term may appear at low fields, while for moderate fields, still in the conical state, the system energy is dominated by nearest-neighbor pairwise and equilateral three-body terms, with the latter providing a surprisingly large fraction of the total binding energy. This emphasizes that theoretical descriptions of skyrmion crystals, skyrmion dynamics and manipulation of skyrmions in helimagnets must go beyond purely pairwise models to realistically capture the properties of interest in experiment and applications.

\section*{Acknowledgments}

This work was supported by the Research Foundation - Flanders (FWO-Vlaanderen) and Brazilian agencies FAPESP, FACEPE, CAPES and CNPq.
N.P.V. acknowledges
funding from
Funda\c{c}\~{a}o de Amparo \`{a} Pesquisa do Estado de S\~{a}o Paulo - FAPESP (Grant 2024/13248-5).
J.C.B.S. acknowledges
funding from
Funda\c{c}\~{a}o de Amparo \`{a} Pesquisa do Estado de S\~{a}o Paulo - FAPESP (Grant 2021/04941-0).
We would like to thank Prof. Felipe F. Fanchini and FAPESP for providing part of the computational resources used in this
work (Grants: 2021/04655-8 and 2024/02941-1, respectively). The other part of the computational resources used in this work was provided by the VSC (Flemish Supercomputer Center), funded by Research Foundation-Flanders (FWO) and the Flemish Government -- department EWI.

\bibliography{mybib}
\end{document}